\pdfoutput=1
\documentclass[10pt,twocolumn]{article}
\usepackage[letterpaper,margin=0.8in,columnsep=0.3in]{geometry}
\usepackage[T1]{fontenc}
\usepackage[nofontspec,scaled=.98]{XCharter}
\usepackage[charter,vvarbb,scaled=1.05]{newtxmath}
\usepackage[scaled=0.92]{inconsolata}
\usepackage[protrusion=true,expansion=false]{microtype}
\usepackage{graphicx}
\usepackage[font=small,labelfont=bf,labelsep=period,skip=5pt]{caption}
\usepackage{titlesec}
\usepackage[colorlinks=true,allcolors=black,bookmarksnumbered=true]{hyperref}
\usepackage{orcidlink}
\titleformat{\section}{\normalfont\large\bfseries}{\thesection}{0.6em}{}
\titlespacing*{\section}{0pt}{10pt plus 2pt minus 2pt}{4pt}
\titleformat{\paragraph}[runin]{\normalfont\normalsize\bfseries}{}{0pt}{}[.]
\titlespacing*{\paragraph}{0pt}{5pt plus 1pt minus 1pt}{0.5em}
\newcommand{\HeldTraces}{57}
\newcommand{\HeldVolumes}{38}
\newcommand{\EstMedLo}{0.46}
\newcommand{\EstMedHi}{0.52}
\newcommand{\TableMed}{0.68}
\newcommand{\TableAhead}{50}

\newcommand{\TableEighty}{16}
\newcommand{\OneCellMed}{0.60}

\newcommand{\CallsSlow}{52}
\newcommand{\ReadsSlow}{30}
\newcommand{\SlowCallShare}{73}
\newcommand{\PeriodSlow}{53}

\newcommand{\EuclidFlagGap}{4}

\newcommand{\AdmTable}{0.61}
\newcommand{\AdmEstLo}{0.47}
\newcommand{\AdmEstHi}{0.55}

\newcommand{\PerTable}{0.52}

\newcommand{\PerEstHi}{0.62}
\newcommand{\PerEuclid}{0.47}
\newcommand{\PerMahal}{0.51}
\newcommand{\DevWrongLo}{27}
\newcommand{\DevWrongHi}{67}

\newcommand{\DevWrongPerLo}{29}
\newcommand{\DevWrongPerHi}{43}

\newcommand{\ArtAccLo}{81}
\newcommand{\ArtAccHi}{90}
\newcommand{\ArtAccDiff}{0.5}
\newcommand{\FilteredLo}{5}
\newcommand{\FilteredHi}{18}

\newcommand{\TableNs}{21}

\newcommand{\EuclidNsWhole}{19}

\newcommand{\HeimdallUs}{1.7}

\newcommand{\DirHeimAhead}{47}
\newcommand{\DirTableMed}{0.63}
\newcommand{\DirHeimMed}{0.55}
\newcommand{\DirDevHeimMed}{0.43}
\newcommand{\DirDevTableMed}{0.83}

\newcommand{\WrongHeim}{42}

\newcommand{\ArtRatioLo}{1.4}
\newcommand{\ArtRatioHi}{2.9}

\newcommand{\IdRoundsSix}{five}

\newcommand{\IdDistWithin}{9}
\newcommand{\IdOracleGain}{13}

\newcommand{\CalmSlow}{2}
\newcommand{\CalmCalledSlow}{96}

\newcommand{\HeimAccLo}{63}
\newcommand{\HeimAccHi}{82}
\newcommand{\StudyCutLo}{56}
\newcommand{\StudyCutHi}{57}

\newcommand{\IdCutLo}{3}
\newcommand{\IdCutHi}{24}

\newcommand{\T}{\ensuremath{T}}
\newcommand{\us}{\,\textmu s}
\newcommand{\ns}{\,ns}
\hypersetup{pdftitle={Predicting, Fast and Slow: SSD Already Knows Which Guess Is Wrong},pdfauthor={Philipp Bogdan}}
\begin{document}
\twocolumn[{%
\centering
{\LARGE\bfseries Predicting, Fast and Slow:\\[3pt] SSD Already Knows Which Guess Is Wrong\par}
\vspace{12pt}
{\large Philipp Bogdan\,\orcidlink{0009-0009-7405-3555}\par}
\vspace{3pt}
{\small Imperial College London\par}
{\small\href{mailto:philipp.bogdan25@imperial.ac.uk}{\nolinkurl{philipp.bogdan25@imperial.ac.uk}}\par}
{\small November 2026\par}
\vspace{16pt}
\begin{minipage}{0.86\textwidth}
\textbf{Heimdall, a learned controller, calls slow on \CallsSlow\% of an SSD's reads when \ReadsSlow\% are slow. Uncertainty estimators, the remedy prescribed for its wrong guesses, cannot find them: nine of them do no better than a coin. The drive can.}
On one trace, \CalmSlow\% of the reads of 4 to 8\,KiB that found less than 4\,KiB outstanding while the drive was answering came back slow, and Heimdall called \CalmCalledSlow\% of them slow. A 4,096-entry table of how the drive's recent reads came back, untrained and \TableNs\ns{} a decision, ranks the wrong guesses at a median AUROC of \TableMed{} on \HeldTraces{} held-out traces, where the nine sit at \EstMedLo{} to \EstMedHi, and it predicts slow reads better than Heimdall itself.
\end{minipage}\par
\vspace{14pt}
\includegraphics[width=\textwidth]{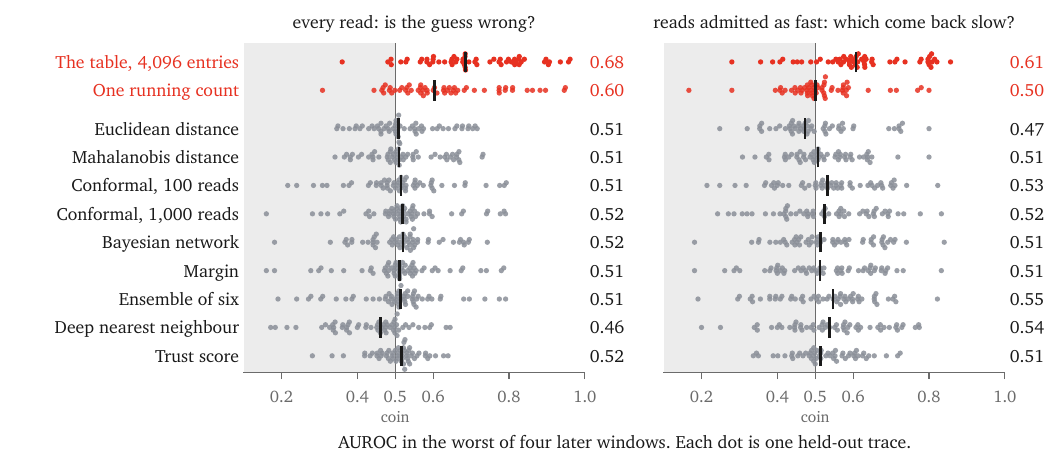}\par
\captionsetup{type=figure}
\caption{Which of Heimdall's guesses are wrong? Each dot is one of \HeldTraces{} held-out traces, at the AUROC with which a row ranks wrong guesses above right ones. Bars are medians. Right: only reads Heimdall admits as fast, where knowing that it calls slow too often no longer helps. Red rows learn from completed reads. Grey rows are the nine uncertainty estimators.}
\label{fig:heldout}
\vspace{12.8pt}
}]

\section{The estimators flip a coin}
\label{sec:coin}
Heimdall~\cite{heimdall} puts a small neural network in front of each SSD of a replicated pair. Before a read is submitted the network guesses fast or slow, and a slow guess sends the read to the other drive. Gohil et al.~\cite{gohil} add an uncertainty estimator that scores each guess, so that a distrusted read can go to both drives. At microsecond budgets they prescribe distance from the training data: ``a distance-based estimator must be selected, as it is the only viable option''. Their evidence is latency: a \StudyCutLo{} to \StudyCutHi\% cut at p99.9. Whether distance picks out Heimdall's wrong guesses, they did not measure.

\begin{figure*}[t]
\centering
\includegraphics[width=\textwidth]{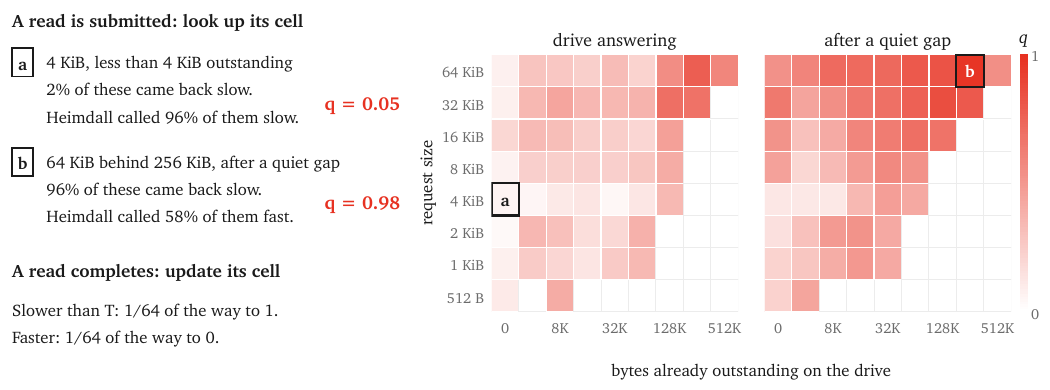}
\caption{The table after three minutes of one held-out trace, an MSR Cambridge media-server volume, for reads submitted with nothing outstanding for longer than \T, the knee of the first minute's read latencies. Sizes and backlogs are classes one power of two wide, named by their lower edge. Blank cells saw fewer than 20 reads. Slow reads are large reads behind a backlog, more so just after the drive has gone quiet. Heimdall's twelve inputs carry neither the backlog in bytes nor the quiet.}
\label{fig:table}
\end{figure*}

We measured it. Heimdall's own pipeline trains the network on the first minute of a three-minute replay, as its long-term evaluation deploys it~\cite[\S7]{heimdall}, and every read of the next two minutes is scored. A read is slow when its latency exceeds the knee of the replay's read-latency distribution, found as LinnOS~\cite{linnos} and Heimdall find it. A guess is wrong when the read comes back on the other side of that knee: the task as Gohil et al. state it, ``binary latency classification (fast/slow) for each I/O''. An estimator's score is the AUROC with which it ranks wrong guesses above right ones, in the worst of four 30-second windows.

Nine estimators compete. Three families are those of Gohil et al., rebuilt from their description: distance, conformal scores and a Bayesian network~\cite{blundell}. Four are standard: the network's margin, an ensemble~\cite{ensembles}, deep nearest neighbours~\cite{knn} and trust score~\cite{trust}.

None finds the wrong guesses (Figure~\ref{fig:heldout}): on \HeldTraces{} held-out traces their median AUROCs run from \EstMedLo{} to \EstMedHi. The traces are the busiest stretches of \HeldVolumes{} volumes, nearly all from the UMass collections~\cite{umass} and the rest from MSR Cambridge~\cite{msr}, at their recorded pace or eight times faster. Each was replayed once on a WD\_BLACK SN770, and nothing here was tuned on them. Used as Gohil et al. use it, to flag the inputs beyond the 90th percentile of the training distances, Euclidean distance says little: a guess it flags is wrong a median \EuclidFlagGap{} points more often than one it passes, on traces where a median \WrongHeim\% of Heimdall's guesses are wrong.

\section{The wrong guesses are a bias}
\label{sec:bias}
Doubt is the wrong thing to look for. Heimdall is biased. Over the \HeldTraces{} traces it calls slow on a median \CallsSlow\% of reads, where a median \ReadsSlow\% come back slow, and a median \SlowCallShare\% of its wrong guesses are slow calls on reads that come back fast. Figure~\ref{fig:table} shows it at its starkest. An estimator that asks how unfamiliar an input is, or how torn the network is, cannot see a bias. One number can: a single running count of how often reads come back slow, blind to the read in front of it, outranks all nine estimators (median \OneCellMed).

The bias is taught. Heimdall trains on period labels, which mark whole stretches of a trace slow: a median \PeriodSlow\% of the scored reads, where a median \ReadsSlow\% come back slow.

None of this is a weak copy of Heimdall. Its artefact reports \ArtAccLo{} to \ArtAccHi\% accuracy on its twelve traces, and our reimplementation of its pipeline reproduces that to within \ArtAccDiff{} points, labels and features identical row for row. But that test filters \FilteredLo{} to \FilteredHi\% of a trace's rows as noise, splits the rest at random into training and test halves, and describes each I/O by the three submitted before it, finished or not. Deployed, the network sees the last three I/Os that completed, and a future it was not trained on. On those traces it is then wrong on \DevWrongPerLo{} to \DevWrongPerHi\% of reads by its own period labels and \DevWrongLo{} to \DevWrongHi\% by each read's latency. Heimdall's long-term evaluation reports \HeimAccLo{} to \HeimAccHi\% accuracy after one training session~\cite[\S7]{heimdall}.

\section{The drive already knows}
\label{sec:table}
The table keeps one number per situation: how often reads submitted like this one have lately come back slow (Figure~\ref{fig:table}). A situation is four things known when a read is submitted: its size, the bytes already outstanding on the drive, the age of the oldest outstanding request, and whether the drive has just been quiet. With 32, 16, 4 and 2 classes of each, that makes 4,096 entries, each starting at $1/2$. When a read completes, the entry $q$ it was submitted under moves $1/64$ of the way to 1 if the read took longer than \T, and to 0 otherwise:
\[ q \leftarrow q + \bigl(\mathbf{1}\{\ell>\T\}-q\bigr)/64 . \]
\T{} is that knee taken from the first minute alone, without hindsight, so the table needs no training and no label but each read's own latency. A fast guess is suspect in proportion to $q$, a slow guess to $1-q$. The table keeps learning from every completion. The network and the nine estimators are fitted to the first minute and then frozen. None of the machinery is new: branch predictors have kept such confidence counters since 1996~\cite{jacobsen}.

The table was shaped on traces from Heimdall's artefact, and no held-out trace took part in its design. On the \HeldTraces{} held-out traces it ranks Heimdall's wrong guesses at a median AUROC of \TableMed{} and beats the best of the nine, picked per trace with hindsight, on \TableAhead{} of the \HeldTraces. The level is modest: \TableEighty{} traces reach 0.80. And most of the lead is the bias. Among reads Heimdall admits as fast, where knowing the bias no longer helps, the table scores \AdmTable{} against \AdmEstLo{} to \AdmEstHi{} for the nine, level with the best of them picked in hindsight.

Judged by Heimdall's period labels instead of each read's latency, the table is a coin (\PerTable) and the best of the nine reaches \PerEstHi: the table learns what reads did, not what the labeller says. The two distances fail under both (\PerEuclid{} and \PerMahal). Nor does the lead carry to two other hedging decisions~\cite{alibaba,thesios}, with controllers we trained on each request's own outcome: there the table trails the network's margin on one and leads it narrowly on the other.

It costs what distance costs. Timed alone on one core of an Intel i9-13900KF, the table takes \TableNs\ns{} a decision and Euclidean distance \EuclidNsWhole\ns, against \HeimdallUs\us{} for a fixed-point copy of Heimdall's inference. The microsecond budget that left distance as ``the only viable option'' admits the drive's own outcomes too.

And it predicts slow reads better than the network it checks. Take $q$ itself as the forecast that a read will be slow: its median AUROC is \DirTableMed{} against \DirHeimMed{} for Heimdall's output, and it is ahead on \DirHeimAhead{} of the \HeldTraces{} traces. On the artefact's own twelve traces Heimdall's output has a median AUROC of \DirDevHeimMed, below a coin, and the table \DirDevTableMed. The fault is not the architecture: trained on each read's own latency, with the inputs it sees when deployed, the same network draws level with the table. The table gets there untrained.

\section{The tail cannot referee}
\label{sec:tail}
\begin{figure}[!b]
\centering
\includegraphics[width=0.9\columnwidth]{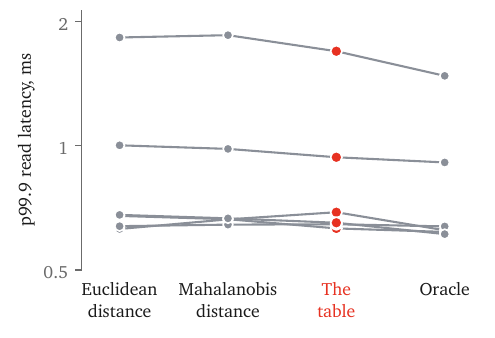}
\caption{Whoever picks the reads, the tail barely moves. Each line is one of six trace pairs on two identical NVMe drives, with hedging on: its p99.9 read latency, the median over rounds, when the guesses to distrust are chosen by a distance, by the table, or by an oracle told which were wrong in an earlier replay.}
\label{fig:tail}
\end{figure}
The evidence offered for the estimators was latency, above all the tail, and the tail does not care who picks the reads (Figure~\ref{fig:tail}). We replayed the six trace pairs of Heimdall's artefact on two identical enterprise NVMe drives, for \IdRoundsSix{} rounds, keeping only replays during which the drives were otherwise idle. The comparison is made with hedging on~\cite{tail}: a read still unanswered after the 98th percentile of the first minute's read latencies gets a copy on the other drive. At the median round, the better of the two distances leaves p99.9 within \IdDistWithin\% of where the table leaves it, on every pair, and even an oracle, told which guesses were wrong in an earlier replay, gains at most \IdOracleGain\% on whichever of them does best.

Nor is the reported cut reproduced. Rebuilt from their description and run without hedging, the better of the two distances cuts Heimdall's p99.9 by \IdCutLo{} to \IdCutHi\% on these drives, pair by pair at the median round, where Gohil et al. report \StudyCutLo{} and \StudyCutHi\% on theirs. We cannot say which difference of drives, traces or training accounts for the gap.

And part of any cut needs no model. In the replay logs that Heimdall's artefact ships, its own hedging baseline, with no model, has the lower p99.9 on all six pairs, by \ArtRatioLo{} to \ArtRatioHi{} times, and Heimdall the lower mean on all six. Heimdall's paper says as much: hedging ``delivers shorter tail latencies above p99''~\cite{heimdall}.

\paragraph{What to take away}
Heimdall's wrong guesses are a bias, and doubt cannot see a bias. Before bolting an uncertainty estimator onto a learned controller, check that it finds the wrong guesses, read by read, because the tail will not say. And ask the device first. Every completed read says how it came back, and a 16\,KiB table of those answers, at \TableNs\ns{} a decision, knows more about Heimdall's next guess than nine estimators do, and more about the next read than Heimdall does.

\paragraph{Supplement}
One script recomputes every number here and every measurement left out:\\
\href{https://github.com/philippbogdan/predicting-fast-and-slow}{\nolinkurl{github.com/philippbogdan/predicting-fast-and-slow}}


{\footnotesize
\begin{thebibliography}{99}
\setlength{\itemsep}{0pt}\setlength{\parskip}{0pt}
\bibitem{heimdall} D.~H. Kurniawan, R.~A. Putri, P.~Qin, K.~S. Zulkifli, R.~A.~O. Sinurat, J.~Bhimani, S.~Madireddy, A.~I. Kistijantoro, and H.~S. Gunawi. Heimdall: Optimizing storage I/O admission with extensive machine learning pipeline. In \emph{EuroSys}, 2025.
\bibitem{gohil} V.~Gohil, N.~Stojkovic, N.~Bashir, S.~Dev, G.~Upasani, D.~Lo, P.~Ranganathan, and C.~Delimitrou. When machine learning isn't sure: Building resilient ML-based computer systems by embracing uncertainty. In \emph{MLSys}, 2026.
\bibitem{tail} J.~Dean and L.~A. Barroso. The tail at scale. \emph{Communications of the ACM}, 56(2), 2013.
\bibitem{jacobsen} E.~Jacobsen, E.~Rotenberg, and J.~E. Smith. Assigning confidence to conditional branch predictions. In \emph{MICRO}, 1996.
\bibitem{linnos} M.~Hao, L.~Toksoz, N.~Li, E.~E. Halim, H.~Hoffmann, and H.~S. Gunawi. LinnOS: Predictability on unpredictable flash storage with a light neural network. In \emph{OSDI}, 2020.
\bibitem{blundell} C.~Blundell, J.~Cornebise, K.~Kavukcuoglu, and D.~Wierstra. Weight uncertainty in neural network. In \emph{ICML}, 2015.
\bibitem{ensembles} B.~Lakshminarayanan, A.~Pritzel, and C.~Blundell. Simple and scalable predictive uncertainty estimation using deep ensembles. In \emph{NeurIPS}, 2017.
\bibitem{knn} Y.~Sun, Y.~Ming, X.~Zhu, and Y.~Li. Out-of-distribution detection with deep nearest neighbors. In \emph{ICML}, 2022.
\bibitem{trust} H.~Jiang, B.~Kim, M.~Y. Guan, and M.~Gupta. To trust or not to trust a classifier. In \emph{NeurIPS}, 2018.
\bibitem{umass} UMass Trace Repository. Storage traces: Financial and WebSearch. \url{https://traces.cs.umass.edu/docs/traces/storage/}.
\bibitem{msr} D.~Narayanan, A.~Donnelly, and A.~Rowstron. Write off-loading: Practical power management for enterprise storage. In \emph{FAST}, 2008.
\bibitem{alibaba} S.~Luo, H.~Xu, K.~Ye, G.~Xu, L.~Zhang, G.~Yang, and C.~Xu. The power of prediction: Microservice auto scaling via workload learning. In \emph{SoCC}, 2022.
\bibitem{thesios} P.~M. Phothilimthana, S.~Kadekodi, S.~Ghodrati, S.~Moon, and M.~Maas. Thesios: Synthesizing accurate counterfactual I/O traces from I/O samples. In \emph{ASPLOS}, 2024.
\end{thebibliography}
}
\end{document}